\documentclass[aps]{revtex4}
\usepackage{amsmath,amssymb,dcolumn}
\usepackage{epsfig}
\newif\ifdviout
\dvioutfalse
\def\FigDir{figures}

\usepackage{amsmath}
\usepackage{amssymb}
\usepackage{mathrsfs}

\def\defscript{\mathscr}

\def\H{{\defscript H}}

\def\ifempty#1{\def\tmpdata{#1}\ifx\tmpdata\empty }

\def\linebreak{\hfill\break}

\def\bra<#1|{\langle #1\rvert}
\def\ket|#1>{\lvert#1 \rangle}
\def\braket<#1|#2>{\langle #1|#2 \rangle}

\def\otop#1{\hbox{$#1\kern-0.1em$\llap{\hbox{\raise1.7ex\hbox{$\scriptstyle\circ$}}}} }

\def\inpare#1{\left(#1\right)}
\def\bigpare(#1){\left(#1\right)}
\def\inrbra#1{\left\{ #1 \right\}}
\def\insbra#1{\left[ #1 \right]}

\def\bigbra[#1]{\left[ #1 \right]}

\def\h{\hat }
\def\t{\tilde }

\def\tend{\rightarrow}

\def\equivalent{\quad\Leftrightarrow\quad}
\def\therefore{\mbox{\setbox0=\hbox{X}\hbox{$\ldotp$}\raise0.7\ht0\hbox{$\ldotp$}\hbox{$\ldotp$}} \quad }
\def\because{\mbox{\setbox0=\hbox{X}\raise0.7\ht0\hbox{$\ldotp$}\hbox{$\ldotp$}\raise0.7\ht0\hbox{$\ldotp$}}\kern0pt }

\def\upin{\hbox{\setbox0=\hbox{$\cup$} \vrule width 0.05 \wd0 height \ht0 depth 0pt \kern - 0.5\wd0 \box0 }}
\def\Frac(#1/#2){\left(\frac{#1}{#2}\right)}

\def\Im{{\rm Im\,}}
\def\Re{{\rm Re\,}}

\def\sdprod{\mathrel{{\setbox0=\hbox{$\displaystyle\times$}\lower0.3\wd0\hbox{$\stackrel{\box0}{\scriptstyle\sim}$}}}}

\def\tosigma#1,{%
    \ifx\tmpindex\relax \def\tmpindex{#1} \let\next=\tosigma
    \else \ifnum\tmpindex=0 1 \else \sigma_\tmpindex \fi
          \ifx#1\relax  \let\next=\relax
          \else \otimes \let\next=\tosigma \def\tmpindex{#1} \fi
    \fi \next}
\def\tspb(#1){\let\tmpindex=\relax\tosigma#1,\relax,}
\def\pd{\partial}

\def\THB{{\mathbb T}}

\def\SHB{{\mathbb S}}

\def\Eq#1{\begin{equation} #1 \end{equation}}
\def\Eqn#1{\Eq{#1 \nonumber}}
\def\Eqr#1{\begin{eqnarray} #1 \end{eqnarray}}

\def\Eqrsub#1{\begin{subequations}\Eqr{#1}\end{subequations}}
\def\Eqrsubl#1#2{\begin{subequations}
  \expandafter\ifx\csname Rlabel\endcsname \relax \label{#1}
  \else \Rlabel{#1} \fi \Eqr{#2}\end{subequations}}
\def\Bitm{\begin{itemize}}
\def\Eitm{\end{itemize}}
\def\Blist#1#2{\begin{list}{#1}{\parsep=0pt \itemsep=0pt%
  \listparindent=0pt #2}}
\def\Elist{\end{list}}
\long\def\ignore#1#2{\def\ignoreflag{#1}\long\def\tmptext{#2}
  \ifnum\ignoreflag>1 #2 \fi}

\def\imo{i}

\def\THB{{\mathbb T}}

\def\SHB{{\mathbb S}}

\begin{document}

\title{Gravitational instability of simply rotating AdS black holes in higher dimensions}

\author{Hideo Kodama}\email{Hideo.Kodama@kek.jp}
\affiliation{Cosmophysics Group, IPNS, KEK and the Graduate University of Advanced
Studies, 1-1 Oho, Tsukuba 305-0801, Japan}

\author{R. A. Konoplya}\email{konoplya_roma@yahoo.com}
\affiliation{Department of Physics, Kyoto University, Kyoto 606-8501, Japan}

\author{Alexander Zhidenko}\email{zhidenko@fma.if.usp.br}
\affiliation{Instituto de F\'{\i}sica, Universidade de S\~{a}o Paulo \\
C.P. 66318, 05315-970, S\~{a}o Paulo-SP, Brazil}

\begin{abstract}
We study the stability of AdS black hole holes rotating in a single two plane for tensor-type gravitational perturbations in $D > 6$ space-time dimensions. First, by an analytic method, we show that there exists no unstable mode when the magnitude $a$ of the angular momentum is smaller than $r_h^2/R$ where $r_h$  is the horizon radius, and $R$ is the AdS curvature radius. Then, by numerical calculations of quasinormal modes, using the separability of the relevant perturbation equations, we show that an instability occurs for rapidly rotating black holes with $a>r_h^2/R$, although the growth rate is tiny (of order $10^{-12}$ of the inverse horizon radius). We give numerical evidences indicating that this instability is caused by superradiance.
\end{abstract}

\pacs{04.30.Nk,04.50.+h}
\maketitle

%T1>Introduction
\section{Introduction}

Recent years, the perturbative study of higher-dimensional black holes became an active research area \cite{Kanti:2008eq,Kodama:2007ph} along with the developments in string theory requiring extra dimensions and various higher-dimensional world models \cite{Horava:1995qa,ArkaniHamed:1998rs,Randall:1999ee}. 

In such a world model, the black hole stability is a crucial problem because no simple uniqueness theorem holds in spacetimes with dimensions higher than four. Instead, it has turned out that a great variety of higher-dimensional objects can exist in higher dimensions: Myers-Perry (MP) black holes, Kaluza-Klein black holes, different types of black strings, branes, rings and saturns \cite{Emparan:2008eg}. In order to know which of these solutions can be realized in nature, we need to test their stability against small metric perturbations. The black hole stability is also conjectured to provide useful information on the phase structure of some conformal field theories through the AdS/CFT correspondence \cite{Maldacena.J1998,Gubser:2000mm,Konoplya:2008rq}.

The systematic investigation of the black hole stability in higher dimensions started from the papers \cite{Ishibashi:2003ap,Kodama:2003jz,Kodama:2003kk}, where the master wave equations for gravitational perturbations of Tangherlini-like static black holes allowing for the cosmological constant $\Lambda$ were derived. Then, with the help of them, the stability of higher-dimensional Schwarzschild black holes was proven analytically \cite{Kodama:2003jz}. The analytical tools used there, however, did not work well in the presence of the non-vanishing $\Lambda$-term or electromagnetic charge. Therefore, the stability problem in these cases have been studied numerically. For example, the stability of D-dimensional Schwarzschild-de Sitter black holes was shown numerically in \cite{Konoplya:2007jv}, and that of Reissner-Nordstr\"om (RN) black holes in \cite{Konoplya:2008au}. Further, recently, a gravitational instability has been found for RNdS black holes in $D \geq 7$ space-time dimensions, when both charge and $\Lambda$-term are large enough \cite{Konoplya:2008au}.

When we extend the analysis to the system with negative $\Lambda$ and non-vanishing charge, a subtlety arises. It is because the Reissner-Nordstr\"om-anti-de Sitter (RNAdS) metric can be a solution in the D-dimensional Einstein-Maxwell theory and in the $\mathcal{N} =8$ gauged supergravity theory at the same time. The latter theory includes a dilaton. Thus, the dynamics of black hole perturbations evidently depends on the theory in which we consider the RNAdS metric. For example, the RNAdS black holes are dynamically unstable in the $\mathcal{N} =8$ gauged supergravity with $D=4$ when the AdS black hole is large and highly charged \cite{Gubser:2000mm}, while the RNAdS black holes in the pure Einstein-Maxwell theory are dynamically stable \cite{Konoplya:2008rq}.

Motivated by the low-energy limit of string theory, D-dimensional black holes in the Gauss-Bonnet theory were also studied and proved to be unstable for large values of the Gauss-Bonnet coupling and only for $D =5,6$ space-time dimensions \cite{Beroiz:2007gp,Konoplya:2008ix2}. The stability and quasinormal modes of Kaluza-Klein black holes with squashed horizon was considered in \cite{Ishihara:2008re}, while the evolution of instability in time domain for  black strings was analyzed in \cite{Konoplya:2008yy}.

Thus, the stability of higher-dimensional static black holes has been studied quite well up to now (though not completely). Unlike the static case, the stability of rotating black holes in higher dimensions has not been proven so far, because the perturbation equations for them cannot be reduced to a set of decoupled wave-like equations and the separation of variables is a hard problem as well. Nevertheless, many studies suggest that rotating black holes tend to be more unstable compared with static ones. For example, when we consider a bosonic field around a rotating black hole, there appear modes with frequencies $\Re \omega < m \Omega_h$ that extract a part of the rotational energy of the black hole at scattering.  For such a mode, a reflected wave has a larger amplitude than an incident one. This effect of amplification of the incident wave was called the superradiance and calculated for the first time for Kerr black holes by Starobinsky \cite{Starob}. Because the effective potential for a massive field has a local minimum, the wave amplified by superscattering can be reflected back onto the black hole again by the wall around the minimum. The repetition of this superradiance and reflection can lead to the growth of the wave amplitude  without bound for a massive field around a Kerr black hole \cite{super}, although  its  growth rate is in most cases too small to have significant effects in realistic situations.

This type of instability is also expected to occur for rotating AdS black holes because the asymptotic behavior of the AdS spacetime produces a diverging effective potential barrier at infinity even for massless fields \cite{Cardoso:2006wa}. In fact, rotating AdS black holes with all equal angular momentum components in spacetimes of odd dimensions with $D \geq 7$ were shown to be unstable for a special class of tensor-type gravitational perturbations in  \cite{Kunduri:2006qa}.
 
The main purpose of the present paper is to investigate the possibility of this type of instability for another special class of black holes that are much more physically motivated: the asymptotically AdS simply rotating black hole, which   is a black hole with only one non-vanishing angular momentum component. Indeed, if black holes are created by particle collisions, the colliding particles, being located on our four-dimensional brane, will create a black hole rotating in a two plane contained in the brane dimensions. One specific feature of such a simply rotating black hole is that the metric has a sphere component and as a consequence, its gravitational perturbations can be classified into scalar, vector and tensor types and expanded by tensor harmonics as in the case of static black holes. In contrast to that case, however, the perturbed Einstein equations cannot be reduced to ordinary differential equations (ODEs) by this harmonic expansion even after the Fourier transformation with respect to time and the rotational angular coordinate.  Nevertheless, the problem can be reduced to a single two-dimensional partial differential equation (PDE) for tensor-type gravitational perturbations, which turns out to be identical to the massless free scalar field equation on the black hole background and can be reduced to ODEs by the separation of variables.

In the present paper, we use these features to study the stability of simply rotating higher-dimensional AdS black holes for tensor-type gravitational perturbations. Because the tensor-type perturbations on $S^n$ exist only for $n\ge 3$,  the analysis applies only to the spacetime dimension $D=n+4 \geq 7$. We investigate the stability by two methods. One is an analytic method motivated by the general arguments of Hawking and Reall \cite{Hawking.S&Reall1999}.  By this method, we prove that the simply rotating black hole is stable against tensor-type gravitational perturbations in the slow rotation regime  $a<r_h^2/R$, where $r_h$ is the horizon radius, and $R$ is the AdS curvature radius. This method, however, cannot be used to find an instability. Hence, we perform a detailed numerical analysis of the quasi-normal frequencies for tensor-type perturbations.  We will show that in addition to the decaying part of the spectrum corresponding to damped quasinormal modes, there are unstable modes that can be interpreted as being produced by superradiance. Indeed, we numerically confirm with very good accuracy that superradiant modes are unstable and that non-superradiant modes are stable in all the cases we calculated.

The paper is organized as follows. Section II introduces the basic features of the background metric for the simply rotating black hole. In Sec. III, we give the tensor-type perturbation equations for this black hole and show that it can be reduced to two weakly coupled  wave-like equations. In Sec IV, the stability in the slow rotating case is proved analytically with the help of an energy integral. Section V explains the numerical methods to determine the quasi-normal frequencies by solving the weakly coupled mode equations for the angular part and the radial part. Section VI discusses the numerical results obtained for damped quasi-normal and unstable superradiant modes. In Sec. VII, we summarize the results and comment on open questions.

%T1>BG sol
\section{Background Solution}

In this section, we summarize the basic features of the simply rotating asymptotically AdS black hole solution in higher dimensions that is used as a background for perturbations.

%T2>Metric
\subsection{Metric}

The general higher-dimensional rotating black hole solution with non-vanishing cosmological constant found by Gibbons, Lu, Page and Pope \cite{Gibbons.G&&2005} has $N$-independent angular-momentum parameters $a_1,\cdots,a_{N}$, in addition to the mass parameter $M$, where $N=[(D-1)/2]$ for $D$-dimensional spacetimes. In the asymptotically flat case, i.e., for the Myers-Perry solution, these parameters correspond to the eigenvalues of the antisymmetric matrix $J_{ab}$ representing the total angular momentum, and each of them to the rotation in one of the $N$ mutually orthogonal 2-planes \cite{Myers.R&Perry1986}.

Although the expression for the metric for generic momentum is rather complicated, it becomes quite simple for the simply rotating case in which the black hole rotates only in a single two plane \cite{Hawking.S&Hunter&Taylor-Robinson1999}:
\Eq{
a_1=a,\ a_2=\cdots=a_{N}=0.
}
The metric for the higher-dimensional AdS-Kerr black hole in this case can be written in the Boyer-Lindquist coordinates as
\Eqr{
ds^2 &=& \frac{X}{X_0^2\rho^2}\insbra{\frac{2M}{r^{D-5}}X -X_0(1-\lambda r^2)\rho^2} dt^2
% \notag\\&&
   -\frac{4aMX\sin^2\theta}{X_0^2\rho^2 r^{D-5}}dtd\phi
% \notag\\   &&
+\frac{\sin^2\theta}{X_0^2\rho^2}\insbra{X_0(r^2+a^2)\rho^2+\frac{2a^2M}{r^{D-5}}\sin^2\theta}
   d\phi^2
  \notag\\&&
   + \frac{\rho^2}{\Delta}dr^2+ \frac{\rho^2}{X}d\theta^2
   +r^2\cos^2\theta d\Omega_{D-4}^2,
\label{KadS:metric:BL}
}
where $d\Omega_n^2$ is the metric of the $n$-dimensional unit sphere, and
\Eqr{
&&\Delta := (r^2+a^2)(1-\lambda r^2) - \frac{2M}{r^{D-5}},\\
&& \rho^2:= r^2 + a^2\cos^2\theta,\quad
   X=1+\lambda a^2 \cos^2\theta,\quad
   X_0=1+\lambda a^2.
}

This metric can be written as the direct sum of the metrics of a 4-dimensional base space and spherical fibres as
\Eq{
ds^2= g_{ab}(y)dy^a dy^b + S(y)^2 d\Omega_{D-4}^2,
\label{MetricStructure:simplyrotating}
}
where $a,b=t,\phi,r,\theta$ and
\Eq{
S(y)= r\cos(\theta).
}
The determinant of the base metric is given by
\Eq{
\det(g_{ab})^{1/2}=|1+\lambda a^2|^{-1} \rho^2 \sin\theta.
}
From the above or the expression \eqref{KadS:metric:BL}, the metric is singular for $X_0\equiv1+\lambda a^2=0$. However, it is not obvious from these expressions whether $X_0<0$ is allowed or not. In order to see that this parameter range is not physical, let us rewrite the metric in the form
\Eqr{
ds^2 &=& -\frac{\Delta}{\rho^2}\inpare{dt-\frac{a}{C}\sin^2\theta d\phi}^2
% \notag\\
+\frac{X\sin^2\theta}{\rho^2}\inpare{adt-\frac{r^2+a^2}{C}d\phi}^2
%   \notag\\&&
  + \frac{\rho^2}{\Delta}dr^2+ \frac{\rho^2}{X}d\theta^2
   +r^2\cos^2\theta d\Omega_{D-4}^2.
}
If $X_0<0$, the function $X$ changes the sign when $\theta$ varies from $0$ to $\pi/2$. However, from this expression, it follows that the metric has the wrong signature $[---+\cdots+]$ for $X<0$. Hence, the metric \eqref{KadS:metric:BL} is regular for all values of $\theta$ only when $X_0>0$, i.e., $a^2<-1/\lambda$.

%T2>horizon
\subsection{Horizon}

From
\Eq{
g_{t\phi}^2-g_{tt}g_{\phi\phi}
  =\Delta \frac{X}{X_0^2} \sin^2\theta,
}
the location of the Killing horizon can be determined by the equation
\Eq{
\Delta = (r^2+a^2)(1-\lambda r^2) - \frac{2M}{r^{D-5}}=0.
}

Since the spacetime is singular at $r=0$, in the five dimensional case $D=5$, the black hole has a regular horizon only when
\Eq{
a^2 < 2M.
}
Hence, there exists an upper bound on the angular momentum parameter $a$, as in the case of four dimensions.
In contrast, for $D>5$, $\Delta=0$ has always a single positive root if $\lambda\le0$ and $M>0$.

%T3>Angular velocity
The horizon angular velocity $\Omega_h$ is determined by the condition
\Eq{
k=\pd_t + \Omega_h \pd_\phi,\quad k\cdot k=0
\equivalent
\Omega_h^2 g_{\phi\phi}+2\Omega_h g_{\phi t} + g_{tt}=0,
}
as
\Eq{
\Omega_h= \frac{a(1-\lambda r_h^2)}{r_h^2+a^2}
 = \frac{2a M}{(r_h^2+a^2)^2 r_h^{D-5}}.
\label{Omega_h}
}
%

%T2>Tensor perturb.
\section{Tensor Perturbations}

For spacetime dimensions greater than or equal to five, from the metric structure \eqref{MetricStructure:simplyrotating}, we can decompose gravitational perturbations of the simply rotating Kerr-$\Lambda$ black hole into reduced tensorial types according to the tensorial behavior with respect to $S^{D-4}$ \cite{Kodama.H&Ishibashi&Seto2000,Kodama:2007ph}. In particular, for $D\ge 7$, tensor-type perturbations appear.

%T3>Master eq.
\subsection{Master equation}

In general, tensor-type metric perturbations for the background \eqref{MetricStructure:simplyrotating} can be expanded in terms of a basis of transverse and traceless harmonic tensors on the unit sphere $S^{D-4}$, $\THB_{ij}^{(l,\alpha)}$, as \cite{Kodama:2007ph} 
\Eq{
\delta g_{ab}=0,\quad
\delta g_{ai}=0,\quad
\delta g_{ij}=2 S(y)^2 \sum_{l,\alpha} H_T^{(l,\alpha)}(y) \THB_{ij}^{(l,\alpha)}(z).
}
Here, $i$ and $j$ refer to the coordinates $z=(z^i)$ of the sphere $S^{D-4}$, $l$ labels the eigenvalue of the harmonic tensor $\THB_{ij}^{(l,\alpha)}$ for the Laplace-Beltrami operator $\h\triangle$ on $S^{D-4}$ as
\Eq{
\insbra{\h\triangle + l(l+D-5)-2} \THB_{ij}^{(l,\alpha)}=0,
}
and $\alpha$ is the label to distinguish harmonic tensors with the same eigenvalue.

After this harmonic expansion, the Einstein equations can be reduced to decoupled second-order hyperbolic equations for each amplitude $H_T^{(l,\alpha)}$. Omitting the index $(l,\alpha)$ for simplicity, this master equation reads \cite{Kodama.H&Ishibashi&Seto2000,Kodama:2007ph}  
\Eq{
-\Box H_T -\frac{D-4}{S}g^{ab}\pd_a S \pd_b H_T
  + \frac{l(l+D-5)}{S^2} H_T=0,
}
where $\Box$ denotes the d'Alembertian for the metric $g_{ab}(y)$. Note that this equation is completely identical to the equation obtained from the free massless field equation for a scalar field on the same background by the harmonic expansion with respect to the harmonic functions $\SHB$ satisfying
\Eq{
\insbra{\h\triangle + l(l+D-5)}\SHB=0.
}

After Fourier expanding $H_T$ with respect to $t$ and $\phi$ as
\Eq{
H_T= H(r,\theta) \exp(-i\omega t + im\phi),
}
where $m$ is an arbitrary integer, we obtain
\Eqr{
&& -\Delta \pd_r^2 H
   +\left[\frac{2M}{r^{D-3}}
   -D+2-\frac{(D-4)a^2}{r^2} +\lambda\inrbra{D r^2+(D-2)a^2}
   \right] r \pd_r H
 \notag\\ &&
 -(1+\lambda a^2 \cos^2\theta) \pd_\theta^2 H
%  \notag\\&&
  +\insbra{ \frac{(D-4)\sin^2\theta -\cos^2\theta}{\sin\theta\cos\theta}
   +\lambda a^2\inrbra{D-2-(D-1)\cos^2\theta} \cot\theta } \pd_\theta H
   \notag\\
&& +\left[
   \inrbra{-\frac{a^2}{\Delta}
    \inpare{ (1+\lambda a^2)(1-\lambda r^2)
               -\frac{2\lambda M}{r^{D-5}} }
    +\frac{1+\lambda a^2}{\sin^2\theta} } m^2
    \right.\notag\\
&&\quad
  +\frac{4aM}{r^{D-5}\Delta} m\omega
  +\inpare{ -\frac{(r^2+a^2)^2}{\Delta}
       +\frac{a^2\sin^2\theta}{1+\lambda a^2\cos^2\theta} } \omega^2
  \notag\\
&& \quad \left.
    +l(l+D-5)\inpare{\frac{a^2}{r^2} + \frac{1}{\cos^2\theta} }
    \right] H=0.
}
%

%T3>Separation of variables
\subsection{Separation of variables}

The equation for $H$ can be separated in the following way
\Eq{
H=\t P(r)Q(\theta)
}
to the angular part
\Eqr{
&& (1+\lambda a^2 \cos^2\theta) Q''
  -\inpare{ \frac{(D-4)\sin^2\theta -\cos^2\theta}{\sin\theta\cos\theta}
   +\lambda a^2\inrbra{D-2-(D-1)\cos^2\theta} \cot\theta } Q'
   \notag\\
&& -\inpare{-\mu
    +\frac{1+\lambda a^2}{\sin^2\theta}  m^2
       +\frac{a^2\sin^2\theta}{1+\lambda a^2\cos^2\theta} \omega^2
    +\frac{l(l+D-5)}{\cos^2\theta} } Q=0.
\label{OriginalModeEq:Q}
}
and to the radial part
\Eqr{
&& \Delta \t P''
   +\left[-\frac{2M}{r^{D-3}}
   +D-2+\frac{(D-4)a^2}{r^2} -\lambda\inrbra{Dr^2+(D-2)a^2}
   \right] r \t P'
 \notag\\
&& +\left[ -\mu
   +\frac{a^2 m^2}{\Delta}
    \inpare{ (1+\lambda a^2)(1-\lambda r^2)
               -\frac{2\lambda M}{r^{D-5}} }
    -\frac{4M}{r^{D-5}\Delta}a m\omega
    +\frac{(r^2+a^2)^2\omega^2}{\Delta}
%  \right. \notag\\
%&& \quad \left.
    -\frac{l(l+D-5)a^2}{r^2} \right] \t P=0,
\label{OriginalModeEq:P}
}
where $\mu$ is the separation constant.

Thus, the stability analysis is reduced to the eigenvalue problem of ODEs as in the case of static black holes. In the rotating case, however, there is a subtlety. In contrast to the static case, $\omega^2$ also appears in the ODE for the angular part. Hence, these two equations are coupled as the eigenvalue equation.

%T1>analytic study
\section{Analytic Study}

In this section, we prove the stability of black holes with $\lambda\le-a^2$  against tensor perturbations. The basic idea is to apply Hawking and Reall's general argument \cite{Hawking.S&Reall1999} for the stability of a scalar field in a slowly rotating Kerr-AdS background to the present system that is equivalent to such a system. Although it is sufficient for the stability proof if we show that there exists an everywhere causal Killing vector outside the horizon \cite{Kodama.H2008}, in the present paper, we explicitly construct the energy integral corresponding to that Killing vector, because its structure gives us information on instability in the rapidly rotating case studied numerically in the next section.

%T2>Phi eq
In the stability argument in the present section, we do not utilize the separability of the differential equation for $H(r,\theta)$. Instead, we directly work with the function $\Phi$ of $r$ and $x=\cos(2\theta)$ defined by
\Eq{
H = r^{-D/2+2}(r^2+a^2)^{-1/2} (1+x)^{-(D-5)/4} \Phi(r,x).
}
$\Phi$ obeys the following equation:
\Eqr{
&& -(r^2+a^2) \pd_r\inrbra{\frac{\Delta}{r^2+a^2}\pd_r \Phi}
   -\pd_x\inrbra{2(1-x^2)(2+\lambda a^2+\lambda a^2 x) \pd_x \Phi}
   \notag\\
&&
 +\Big[ -(\omega-m\Omega)^2 \frac{F}{\Delta}
  +m(\omega-m\Omega) \frac{4a^3(1-x)M}{(2+\lambda a^2 + \lambda a^2 x)
   (r^2+a^2)^2 r^{D-5}}
%   \notag\\
%&&\qquad
 +m^2 U_1(r,x)  + U_0(r) \Big] \Phi=0,
\label{SMP:TP:PEQ:Phi}
}
where
\Eqr{
\Omega &:=&  \frac{2aM}{(r^2+a^2)^2 r^{D-5}},\\
F  &:=& \frac{(1+\lambda a^2)(r^2+a^2)
   (2r^2+ a^2+a^2 x) + \frac{2a^2(1-x)M}{r^{D-5}}}
   {2+\lambda a^2 + \lambda a^2 x},\\
U_0 &:=& \frac{(D-2)^2r^4+2(D^2-6D+4)a^2r^2+(D-4)^2a^2}{2(r^2+a^2)^2r^{D-3}} M
\notag\\
   &&+ \frac{3}{4} - \frac{(D-4)(D+2)(1+x)+7(1-x)}{8}\lambda a^2
          -\frac{D(D-2)}{4}\lambda r^2
           \notag\\
   && +\frac{a^2(1+\lambda a^2)}{r^2+a^2}
     +\inpare{l+\frac{D}{2}-2}\inpare{l+\frac{D}{2}-3} \frac{a^2}{r^2}
      +\frac{(2l-5+D)^2}{2(1+x)},\\
U_1 &:=& \frac{(1+\lambda a^2)(2r^2+a^2+a^2x)}{(1-x)(r^2+a^2)}
          -\frac{2a^2 M}{(r^2+a^2)^2 r^{D-5}}
    \notag\\
    && +\frac{4a^4(1-x)M^2}{(2+\lambda a^2+\lambda a^2 x)(r^2+a^2)^4 r^{2D-10}}.
}
%

%T2>Effective potential
\subsection{Effective potential}

In the static background, the eigen-frequency $\omega$ for an unstable mode is purely imaginary, that is, $\omega^2<0$,  because the corresponding differential operator is self-adjoint. In contrast, the superradiance arguments \cite{Kodama.H2008} and the stability analysis for the special Kerr-AdS black holes \cite{Kunduri:2006qa} suggest that an instability mode appears at the complex frequency whose real part satisfies the superradiance condition $\omega-m\Omega_h=0$. This is natural because $\omega_*$ defined by
\Eq{
\omega_* := \omega - m \Omega_h
\label{omega*:def}
}
is the frequency with respect to the advanced time on the future horizon that is regular with respect to the null geodesic generator of the horizon \cite{Kodama.H2008}.

On the basis of this observation, we take $\omega_*$ as an effective frequency. In terms of this, the equation for $\Phi$ can be rewritten as
\Eqr{
&& -(r^2+a^2) \pd_r\inrbra{\frac{\Delta}{r^2+a^2}\pd_r \Phi}
   -\pd_x\inrbra{2(1-x^2)(2+\lambda a^2+\lambda a^2 x) \pd_x \Phi}
   \notag\\
&&
 +\inpare{ -\omega_*^2 \frac{F}{\Delta}
  -2 m a \omega_* G + m^2 \t U_1 + U_0}\Phi=0,
\label{SMP:TP:PEQ:Phi:tomega}
}
where
\Eqr{
G &=& \frac{(1+\lambda a^2)(2r^2+a^2+a^2x)}
      {(1-\lambda r^2)(r^2+a^2)(2+\lambda a^2+\lambda a^2x)}
      \notag\\
   && \times \insbra{ 1-\lambda r^2 + \delta(1+\lambda a^2)(r^2+a^2)
        +\frac{2\delta a^2M}{r^{D-5}}\frac{1-x}{2r^2+a^2+a^2x} },\\
\t U_1 &=& \frac{(1+\lambda a^2)^2(2r^2+a^2+a^2x)^2}
          {(1-x)(r^2+a^2)^2 (2+\lambda a^2+\lambda a^2 x)}
     \notag\\
  &&\times  \left[
     1- \frac{a^2(r^2+a^2)(1-x)\inrbra{ 2\delta(1-\lambda r^2)
     +\delta^2(r^2+a^2)(1+\lambda a^2) }}
     {(1-\lambda r^2)(2r^2+a^2+a^2x)}
   \right. \notag\\
  &&
   +\frac{2a^2M\delta (1-x)}{r^{D-5}(1-\lambda r^2)(2r^2+a^2+a^2x)}
      \Big\{ 2+\delta (r^2+a^2)
      \notag\\
  && \qquad
     \times  \frac{2(1+\lambda a^2)(r^2+a^2)
       -a^2(1-x)(2+\lambda a^2-\lambda r^2)}
       {(1-\lambda r^2)(2r^2+a^2+a^2x)} \Big\}
     \notag\\
   && \quad\left.
     +\inpare{ \frac{2a^2M\delta (1-x)}
            {r^{D-5}(1-\lambda r^2)(2r^2+a^2+a^2x)} }^2
    \right].
}
In these expressions, $\delta$ is a function of $r$ defined by
\Eq{
\delta = (1+\lambda a^2)\frac{r^2+a^2}{r_h^2+a^2}
  \frac{r^2-r_h^2}{\Delta},
}
which is always positive outside the horizon.

%T2>behavior of U1
\subsection{Behavior of the potential}

$G$ is always positive, and $\t U_1$ is also positive at the horizon $r=r_h$:
\Eq{
\t U_1(r_h)=\frac{(1+\lambda a^2)(2r_h^2+a^2+a^2x)^2}
  {(1-x)(a^2+r_h^2)(2+\lambda a^2+\lambda a^2x)}>0.
}
However, for $\lambda<0$, $\t U_1$ at $r\tend\infty$ can become negative as
\Eq{
\t U_1 \approx
 \frac{4(1+\lambda a^2)^2}{2+\lambda a^2+\lambda a^2 x}
 \insbra{
     \frac{1+x}{2(1-x)}-\frac{a^2+\lambda r_h^2}
      {2(-\lambda)(r_h^2+a^2)^2}
    }.
}

Now, we show that $\t U_1$ is non-negative everywhere outside the horizon if this asymptotic value is non-negative. For that purpose, we introduce
the parameter $b$ by
\Eq{
a^2= \frac{b}{|\lambda|},
}
which is in the range
\Eq{
0\le b <1.
}
In terms of $b$, $\t U_1$ can be expressed in the units $r_h=1$ as
\Eqr{
\t U_1 &=& \frac{(1-b)^2 }
        {(1-x)|\lambda|(b+|\lambda|)^2(2-b-bx)r^{D-4} \Delta}
   \notag\\
&& \times\Big[
  b^2 \left\{ \lambda^2(2r^{D-2}-r^{D-4}-1)+(b+1)|\lambda|r(r^{D-3}-1)
  +b(r^{D-4}-r)\right\} (1+x)^2
  \notag\\
&& \quad
  +2b|\lambda|\left\{2\lambda^2 r(r^{D-1}-1)+(b+1)|\lambda|
  (r^{D-3}+r^{D-4}-2r)+2b(r^{D-4}-r)\right\} (1+x)
  \notag\\
&& \quad
   +4\lambda^2 (b|\lambda|+2b+|\lambda|) r(r^{D-3}-1)
  +4\lambda^2 (\lambda^2-b) r(r^{D-1}-1)
    \Big].
}
All terms except for the last one in the above expression are non-negative. Further, for $x=-1$ adn $r\sim \infty$, the last term dominates. Therefore, the necessary and sufficient condition for $\t U_1\ge0$ is expressed in the general units as
\Eq{
\lambda^2 r_h^2-b \ge 0 \equivalent a^2 \le-\lambda r_h^4.
}
%

%T2>energy integral
\subsection{Energy integral}

Suppose that $\t\Phi(t,r,x)$ is a solution to the equation
\Eqr{
&& -(r^2+a^2) \pd_r\inrbra{\frac{\Delta}{r^2+a^2}\pd_r \t\Phi}
   -\pd_x\inrbra{2(1-x^2)(2+\lambda a^2+\lambda a^2 x) \pd_x \t\Phi}
   \notag\\
&&
 + \frac{F}{\Delta}\pd_t^2 \t\Phi
  -2i m a  G \pd_t\t\Phi + (m^2 \t U_1 + U_0)\t\Phi=0.
\label{SMP:TP:PEQ:tPhi}
}
Then, for the energy integral defined by
\Eqr{
\H(\t\Phi) &:=& \int_{r_h}^\infty dr \int_{-1}^1 dx
  \left[ \frac{\Delta}{r^2+a^2} |\pd_r\t\Phi|^2
    +2\frac{(1-x^2)(2+\lambda a^2+\lambda a^2x)}{r^2+a^2}
    |\pd_x \t\Phi|^2
    \right.\notag\\
  && \quad \left.
  +\frac{F}{(r^2+a^2)\Delta} |\pd_t \t\Phi|^2
  +\frac{m^2 \t U_1+U_0}{r^2+a^2}|\t\Phi|^2 \right],
}
we have
\Eqr{
\frac{d}{dt} \H(\t\Phi) &=&
  \int_{r_h}^\infty dr\int_{-1}^1 dx
  \left[ \frac{\Delta}{r^2+a^2} \pd_r\pd_t\t\Phi^* \pd_r\t\Phi
%  \right.\notag\\
%  &&\quad
      +2\frac{(1-x^2)(2+\lambda a^2+\lambda a^2x)}{r^2+a^2}
      \pd_x \pd_t\t\Phi^* \pd_x\t\Phi
 \right.  \notag\\
 && \quad\left.
  +\pd_t\t\Phi^*\inpare{
  \frac{F}{(r^2+a^2)\Delta}\pd_t^2\t\Phi
  +\frac{m^2 \t U_1+U_0}{r^2+a^2} \Phi} \right] + \text{cc}
 \notag\\
 &=& \int_{-1}^1dx
   \insbra{ \frac{\Delta}{r^2+a^2}
    \pd_t\t\Phi^* \pd_r\t\Phi}_{r=\infty}
    \notag\\
 &&  +\int_{r_h}^\infty dr\int_{-1}^1 dx
  \frac{\pd_t\t\Phi^*}{r^2+a^2}
  \left[ -(r^2+a^2)\pd_r\inpare{\frac{\Delta}{r^2+a^2}\pd_r\t\Phi}
  \right.\notag\\
 && \quad \left.
      -\pd_x\inrbra{
      2(1-x^2)(2+\lambda a^2+\lambda a^2x)\pd_x\t\Phi}
%      \notag\\
% && \quad\left.
  +\frac{F}{\Delta}\pd_t^2\t\Phi
  +(m^2 \t U_1+U_0) \Phi  \right] + \text{cc}
\notag\\
&=& \int_{r_h}^\infty dr\int_{-1}^1 dx
    \frac{2i ma G}{r^2+a^2} |\pd_t\t\Phi|^2 + \text{cc}
  \notag\\
&=& 0,
}
where we have assumed that $\t\Phi$ falls off sufficiently rapidly at $r\sim\infty$ so that $\H(\t\Phi)$ is finite.

%T3>fig:paramregion
\begin{figure}
\centerline{
\ifdviout
\includegraphics[height=8cm]{\FigDir/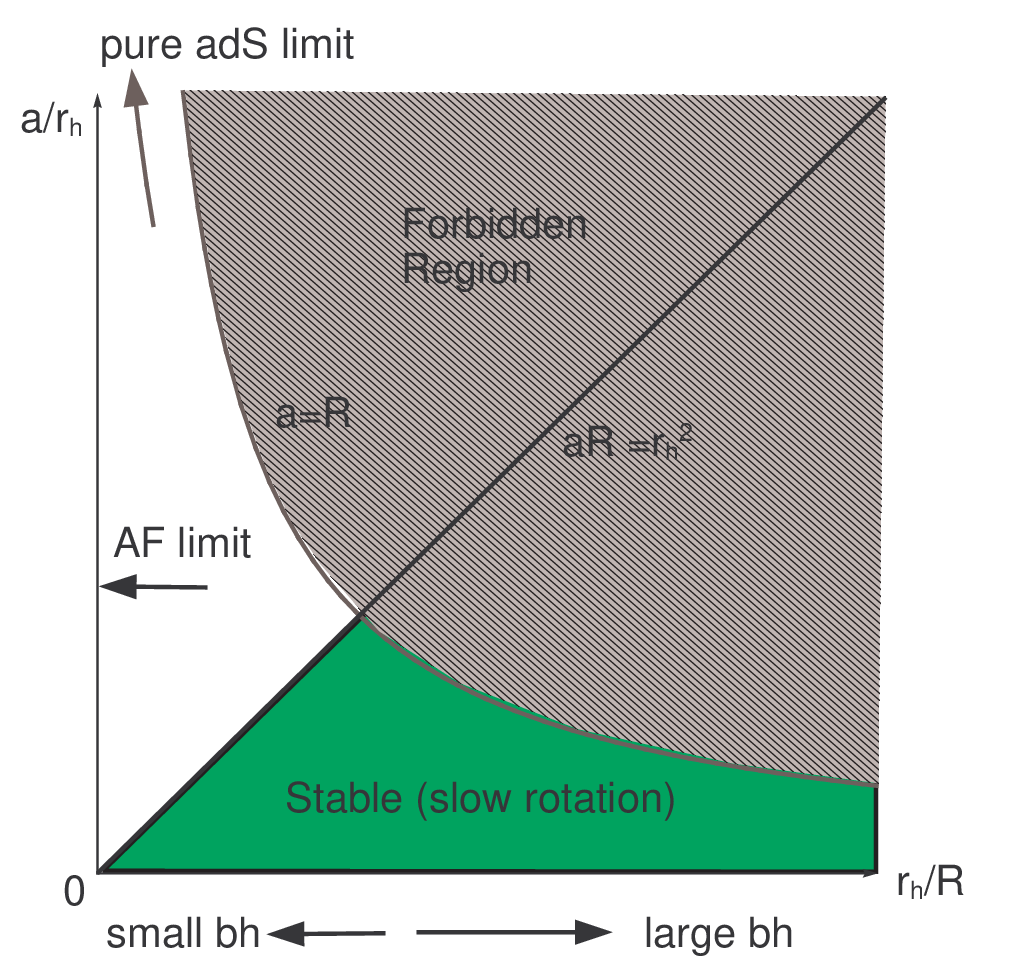}
\else
\includegraphics[height=8cm]{paramregion.eps}
\fi
}
\caption{\label{fig:paramregion} The stable region in the parameter plane for the simply rotating MPadS black hole}
\end{figure}

Hence, $\H(\tilde\Phi)$ is constant in time. From this and $U_0>0$, it follows that if $\t U_1\ge0$ holds, \eqref{SMP:TP:PEQ:tPhi} does not have a normalisable solution of the form $\t\Phi=\Phi(r,x)e^{-i\omega_* t}$ ($\Im\omega_*>0$). In particular, the black hole is stable against tensor perturbations if $\lambda\le -a^2/r_h^4$, or equivalently $a<r_h^2/R$, where $\lambda=-1/R^2$ (see Fig. \ref{fig:paramregion}). Taking into account the regularity condition $a<R$, this condition can be rewritten as $R\Omega_h<1$ because we have
\Eq{
R\Omega_h-1=\frac{(R-a)(aR-r_h^2)}{R(r_h^2+a^2)}
\label{ROmega}}
from \eqref{Omega_h}.

Finally, note that the arguments in this section also imply that the stability of modes do not depend on the sign of $m$ and that instability can occur only for a mode with $m\neq0$.

%T1>Numerical Method
\section{Numerical Method}

%T2>The angular part
\subsection{The angular part: search of $\mu$}

The angular part of the perturbation equation, \eqref{OriginalModeEq:Q}, can be written as
\begin{eqnarray}\label{angular-part}
(1-x^2) \inrbra{2 + a^2(1 + x)\lambda}Q''(x)+\inpare{D-5 - (D-1) x + \frac{a^2}{2} (1 + x) \inrbra{D-3 - (D+1) x} \lambda} Q'(x)&&\\
+\left(\frac{\mu}{2}+\frac{a^2\omega^2(x-1)}{2(2+a^2 (1 + x)\lambda)}+\frac{m^2 (1 + a^2 \lambda)}{x-1}-\frac{l (l + D-5)}{1 + x}\right)Q(x)&=&0\nonumber,
\end{eqnarray}
in terms of the variable $x=\cos(2\theta)$.

The equation (\ref{angular-part}) has four regular singular points: $x=\pm1$, $x=\xi$, and $x=\infty$, where
\Eq{
\frac{a^2}{R^2}=\frac{2}{1+\xi},\quad
\lambda=-\frac{1}{R^2}.
}
If we introduce the rescaled mode function $y(z)$ by
\begin{equation}
Q(x)=(1-x)^{|m|/2}(1+x)^{l/2}(x-\xi)^{\pm R\omega/2}y(z),
\end{equation}
with $x=2z-1$, then this mode equation can be put into the standard form of the Heun's equation\cite{Heun,Maier.R2007},
\begin{equation}
\left(\frac{d^2}{dz^2}+\left[\frac{\gamma}{z}+\frac{\delta}{z-1}+\frac{\epsilon}{z-z_0}\right]\frac{d}{dz}+\frac{\alpha\beta z-q}{z(z-1)(z-z_0)}\right)y(z)=0, \qquad \alpha+\beta+1=\gamma+\delta+\epsilon,
\end{equation}
where
\Eqrsub{
&& \alpha =\frac{1}{2} (l + |m| \pm R\omega), \qquad
   \beta = \frac{1}{2} (l + |m| + D-1 \pm R\omega),\\
&& \gamma=\frac{1}{2} (2 l + D-3), \qquad
   \delta =1+|m|, \qquad
   \epsilon = 1\pm R\omega,\\
&& q=-\frac{m^2}{4}+\frac{(l\pm R\omega)(l+D-3\pm R\omega)}{4}+ \frac{\xi+1}{8}\inrbra{(l+|m|)(l+|m|+D-3) -\mu}, \\
&& z_0 = \frac{\xi+1}{2}=\frac{R^2}{a^2}\,.
}
Note that since $1+\lambda a^2$ must be positive, $z_0>1$. Due to the symmetry of the equation (\ref{angular-part}) with respect to the transformation $R\rightarrow -R$, both signs in the above equations lead to the same eigenvalues.

The boundary condition on $y(z)$ ($0\le z\le1$) can be determined as follows. First, if we put the equation \eqref{angular-part} into the eigenvalue equation form $\mu Q=L Q$ and require that the operator $L$ be self-adjoint when $\omega^2$ is real, we find that the norm of $Q$ should be defined as
\Eq{
N(Q) \propto \int_{-1}^1 dx (1+x)^{(D-5)/2} |Q|^2.
}
Next, it is easy to see that the fundamental solutions to \eqref{angular-part} behave as $Q\approx (x+1)^{l/2},(x+1)^{-(l+D-5)/2}$ around $x=-1$ and as $Q\approx (1-x)^{\pm |m|/2}$ around $x=1$, respectively. Among these solutions, those that behave as $(x+1)^{l/2}$ around $x=-1$ and as $(1-x)^{|m|/2}$ around $x=1$ ($m\neq0$) become normalizable with respect to $N(Q)$. These conditions are identical to the condition that $y(z)$ is regular at $z=0$ and $z=1$.

Now, under this boundary condition, the eigenvalue $\mu$ can be determined as follows. First, following the method of expansion of the Heun's equation proposed in \cite{Suzuki:1998vy},
$$y(z)=\sum_{j=0}^\infty c_j u_j(z), \qquad u_j(z)=F(-j,j+p;\gamma;z)=(-1)^j\frac{\Gamma(2j+p)j!}{\Gamma(j+\gamma)}P^{(p-\gamma,\gamma-1)}_j(2z-1),$$
we find the three-term recurrence relation for the coefficients $c_j$:
$$
c_{j+1}\alpha_j+c_j\beta_j+c_{j-1}\gamma_j=0,
$$
with
\begin{eqnarray}
\alpha_j&=&-
\frac{(j + 1)(j + p - \alpha + 1)(j + p -\beta + 1)(j + \delta)}
{(2j + p + 2)(2j + p + 1)}\,,\\
\beta_j&=&\frac{\epsilon j(j+p)(\gamma-\delta)+[j(j+p)+\alpha\beta][2j(j+p)+\gamma(p-1)]}{(2j + p + 1)(2j + p - 1)}-z_0j(j+p)-q\,,\\
\gamma_j&=&-
\frac{(j + \alpha - 1)(j + \beta - 1)(j + \gamma - 1)(j + p - 1)}
{(2j + p - 2)(2j + p - 1)}\,,
\end{eqnarray}
where we defined the quantity
$$
p = \gamma+\delta -1 = \alpha+ \beta - \epsilon = l+|m|+\frac{D-3}{2}.
$$

For a given $\omega$, eigenvalues $\mu$ can be found through the equation that can be expressed in terms of an infinite continued fraction as
\begin{equation}\label{cont-fraction}
\beta_0=\frac{\alpha_0\gamma_1}{\beta_1-}\frac{\alpha_1\gamma_2}{\beta_2-}\ldots\,.
\end{equation}
Since the continued fraction is convergent \cite{Suzuki:1998vy}, we are able to find the right-hand side of the equation (\ref{cont-fraction}) with any desired precision by limiting the depth of the infinite continued fraction by some large value, always ensuring that an increase in this value does
not change the final results \cite{Leaver:1985ax}. In order to find the $j^{th}$ root we invert (\ref{cont-fraction}) $j$ times
\begin{equation}\label{cont-fraction-inverted}
\beta_j-\frac{\alpha_{j-1}\gamma_j}{\beta_{j-1}-}\frac{\alpha_{j-2}\gamma_{j-1}}{\beta_{j-2}-}\ldots\frac{\alpha_0\gamma_1}{\beta_0}=\frac{\alpha_j\gamma_{j+1}}{\beta_{j+1}-}\frac{\alpha_{j+1}\gamma_{j+2}}{\beta_{j+2}-}\ldots\,.
\end{equation}
We shall use this numerical procedure in order to find $\mu$ for each given $\omega$. The procedure is convergent and therefore accurate
for all $\omega$ and $a$, and does not contain any approximation.

It is also useful to know the non-rotating limit as a start point for numerical search. For that case, 
$a\rightarrow0$, and  the quantities $z_0$ and $q$ are dominant. The equation (\ref{cont-fraction-inverted}) is then reduced to $\beta_j = 0$, which reads
\begin{equation}\label{non-rotation-limit}
q=-z_0 j(j+p), \qquad j=0,1,2,3\ldots\
\end{equation}
This equation can be easily solved with respect to $\mu$ as
\Eq{
\mu = (l+|m|)(l+|m|+D-3) + 4j(j+p)
=(2j+l+|m|)(l+2j+|m|+D-3),
\label{mu:pureAdS}
}
which does not depend on $\omega$ and $R$ in the non-rotation limit.

%T2>Radial part
\subsection{The radial part: the search of $\omega$}

In terms of $P(r)$ defined by
\Eq{
\t P(r)= r^{-D/2+2} (r^2+a^2)^{-1/2} P(r),
}
the radial equation \eqref{OriginalModeEq:P} can be put into the form
\begin{equation}\label{wave-like}
\frac{d^2 P}{dr_\star^2}
+\left[\left(\omega-\frac{2Mam}{(r^2+a^2)^2r^{D-5}}\right)^2-\frac{\Delta(r)}{(r^2+a^2)^2}U(r)\right]P=0,
\end{equation}
where
$$
dr_\star=\frac{r^2+a^2}{\Delta(r)}dr
$$
is the tortoise coordinate, and $U(r)$ is given by
\Eqr{
U &:=& \mu -\frac{a^2 m^2}{(r^2+a^2)^2}
 \inrbra{(r^2+a^2)(1+\lambda a^2)+ \frac{2M}{r^{D-5}} }
 \notag\\
 && + \frac{(D-2)(D-4)}{4}(1-\lambda a^2) -\lambda a^2
   -\frac{D(D-2)}{4}\lambda r^2
   \notag\\
  && +\inpare{l+\frac{D}{2}-2}\inpare{l+\frac{D}{2}-3} \frac{a^2}{r^2}
     +\frac{a^2(1+\lambda a^2)}{r^2+a^2}
     \notag\\
  && +\frac{\inrbra{(D-2)r^2+(D-4)a^2}^2-8a^2 r^2}{2(r^2+a^2)^2}
      \frac{M}{r^{D-3}}.
\label{EffectivePot:radial}
}

Here, note that the background black hole solution has three parameters $M$, $a$ and $\lambda$. From this point, we use the horizon radius $r_h$ of the black hole in place of the black hole mass $M$ that is related to $r_h$ by
\begin{equation}
2M=r_h^{D-5}(r_h^2+a^2)(1-\lambda r_h^2)\,.
\end{equation}
We often use the curvature radius $R$ instead of the cosmological constant $\displaystyle\lambda=-\frac{1}{R^2}$, which should be in the range $a<R$ due to the regularity condition (see Fig. \ref{fig:paramregion}).

Now, let us specify the asymptotic conditions on $P(r)$ in order to determine the frequency $\omega$. In the asymptotically flat case, these conditions for unstable modes and for quasi-normal modes are formally identical and given by the ingoing condition at horizon and the outgoing condition at spatial infinity pretending $\omega$ to be real. The only difference for these two types of modes are just the sign of the imaginary part of the frequency $\omega$. In contrast, for AdS black holes, some subtlety arises concerning the boundary condition at the spatial infinity.

First, at the horizon, we have
\Eqn{
P(r)\sim(r-r_h)^{\pm\imo\omega_\star(r_h^2+a^2)/\Delta'(r_h)},
}
where $\omega_*$ is the shifted frequency defined in \eqref{omega*:def}. Hence, the quasinormal boundary conditions at the horizon are given by
\Eq{
P(r)=(r-r_h)^{-\imo\omega_*(r_h^2+a^2)/\Delta'(r_h)}(Z_0+{\cal O}(r-r_h)).
}
Next, at the spatial infinity, the two linear independent solutions of $P(r)$ are
\begin{equation}
P_1(r)\sim r^{-D/2}, \qquad P_2(r)\sim r^{(D-2)/2},\label{gen-infinity}
\end{equation}
As far as the instability is considered, it is natural to look for a mode normalizable with respect to the norm that makes the equation \eqref{wave-like} self-adjoint. This requirement is equivalent to the condition that $r^{- 1/2}P$ is bounded at infinity. Hence, it is natural to require the Dirichlet-type boundary condition at infinity:
\begin{equation}\label{infinity-BC}
P(r\rightarrow\infty)\propto r^{-D/2}.
\end{equation}
We will also use this condition when we look for quasi-normal modes. Notice that the choice of this specific fall-off is stipulated by the interpretation of QNMs by the AdS/CFT correspondence, and was used previously in \cite{Friess:2006kw,Konoplya:2008rq} for non-rotating asymptotically AdS black holes.

Now, let us explain the numerical procedure used to determine $\omega$ by solving \eqref{wave-like} under these boundary conditions. For convenience, let us introduce the new function
\begin{equation}
y(r)=\left(1-\frac{r_h}{r}\right)^{\imo\omega_*(r_h^2+a^2)/\Delta'(r_h)}P(r).
\end{equation}
If $P(r)$ satisfies the quasi-normal boundary conditions, $y(r)$ is regular at the event horizon. Since the function $y(r)$ satisfies the linear equation, we fix its scale as
\Eq{
y(r_h)=1.
}
After the scale is fixed, we expand $y(r)$ near the event horizon as
\begin{equation}
y(r)=1+\sum_{j=1}^pC_j\left(1-\frac{r_h}{r}\right)^j+o\left(1-\frac{r_h}{r}\right)^p\,.
\end{equation}
This expansion allows to find $y(r)$ and $y'(r)$ in some region $r_h<r<r_i$ with a high precision, if $r_i$ is close enough to $r_h$. Values $y(r_i)$ and $y'(r_i)$ are used as a boundary condition for the equation (\ref{wave-like}). Since $\mu(\omega)$ is found numerically, we solve the equation (\ref{wave-like}) for each $\omega$ using the $NDSolve$ built-in function in \emph{Mathematica} for $r_i\leq r \leq r_f$, where $r_f\gg r_h$.

We have used the following values for the fundamental modes: $p=20$, $r_f=5000r_h$, $r_i/r_h=1+r_h/r_f$, making sure, that the result does not change if $p$ or $r_f$ increases. For higher overtones, larger values of $r_f$ are necessary in order to achieve the convergence of the whole procedure.

In the general case, the behavior of $P(r)$ at infinity is a superposition of the two solutions (\ref{gen-infinity}) for the Dirichlet and non-Dirichlet boundary conditions, $P_D(r)$ and $P_n(r)$:
\begin{equation}\label{fit-function}
P(r)=Z_D P_D(r)+Z_nP_n(r),
\end{equation}
where $P_D(r)$ satisfies the quasi-normal boundary condition (\ref{infinity-BC}). If $\omega$ is the quasi-normal frequency, the corresponding solution must satisfy the boundary conditions (\ref{infinity-BC}) at the spatial infinity and, thereby, $Z_n=0$.

Thus, our numerical procedure is the following. We integrate the equation (\ref{wave-like}) numerically, using the boundary conditions at $r=r_i$, which correspond to the purely in-going wave at the event horizon. At large distance,  we compare the obtained function $\Psi(r_f)$ with (\ref{fit-function}) and find, thereby, the coefficients $Z_D$ and $Z_n$ for any given value of $\omega$. The quasi-normal modes correspond to the roots of the equation
\begin{equation}\label{QNM-equation}
Z_n(\omega)=0.
\end{equation}
Here, note that as $\omega$ is generally complex, $Z_n(\omega)$ is complex as well.

%T1>QNM analysis
\section{Quasinormal mode analysis}

From here and on, we shall imply that
\begin{equation}
\omega = \Re (\omega) + i \Im (\omega),
\end{equation}
so that $\Im (\omega) > 0$ corresponds to an unstable (growing) mode, and $\Im (\omega) < 0$ corresponds to a stable (damped) mode.

%T2>Stable QNM
\subsection{Stable modes}

Let us note that if the effective potential is not positive definite, it does not necessarily mean instability. Thus for scalar-type gravitational perturbations of the $D =5$ Schwarzschild-anti-de Sitter black holes, the effective potential approaches
$-\infty$ at $r^{*} = 0$ (see for instance Fig. 1 in \cite{Konoplya:2008rq}), yet this does not lead to instability
as the quasinormal modes are damped for that case \cite{Friess:2006kw,Konoplya:2008rq}.

The stability is not proved analytically for $R > a > r_h^2/R$. Thus,  $1 > a > r_h^2$ (in the units $R=1$) is the region suspected for instability for nonzero $m$, and we shall search quasinormal modes in this region for different values of the quantum numbers: $l = 2, 3, ...$, $m = 1, 2,... $, $j = 0, 1, 2, ...$. Note that the results for the opposite rotation case with $a \rightarrow - a$ can be obtained just by reversing the sign of $m$ in those for $a>0$.

We shall also distinguish the following three regimes of asymptotically AdS black holes: 1) large black holes with $r_h \gg R$, 2) intermediate black holes
with $r_h \cong R$, and 3) small black holes with $r_h \ll R$ (see Fig. \ref{fig:paramregion}). This classification, although conventional, is important, because the properties of the quasinormal spectrum are totally different for small and large black holes, while an intermediate regime is a kind of transition region between the other two.

For example, for large asymptotically AdS non-rotating black holes, the ratio of the frequency $\omega$ to the horizon radius $r_h$ is roughly independent of $r_h$ for the tensor and vector gravitational quasinormal modes \cite{Konoplya:2008rq}. We are convinced by numerical computations that the same result holds for rotating black holes. Indeed, for instance, for the fundamental mode in the $D=7$, $a=0.7$ case, we obtained the following values
of $\omega/r_h$: $5.10783- 2.62731 i$, $5.07295- 2.6235 i$, $5.05571- 2.62124 i$,  $5.04554- 2.61974 i$ for $r_h = 10, ~ 14, ~18, ~22$, respectively. This indicates the convergence of the ratio $\omega/r_h$ to a constant in the limit of large black holes.

%T3>fig:overtones
\begin{figure}
\ifdviout
\includegraphics[width= 0.5 \linewidth]{\FigDir/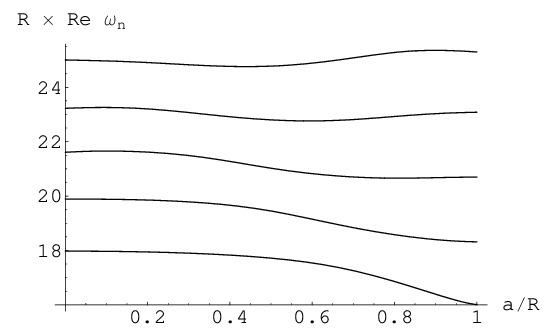},\includegraphics[width= 0.5 \linewidth]{\FigDir/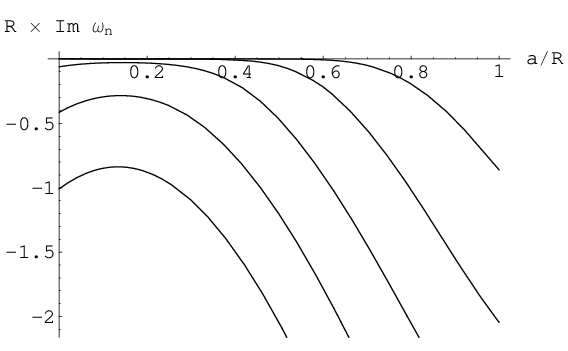}
\else
\includegraphics[width= 0.5 \linewidth]{Re_overtones.eps},\includegraphics[width= 0.5 \linewidth]{Im_overtones.eps}
\fi
\caption{Real (left) and imaginary (right) parts of the first five modes (the fundamental mode and four overtones) for $m=10$, $r_h=0.5$, $D=7$}\label{fig:overtones}
\end{figure}

As $a \rightarrow 0$, and the size of the AdS black holes goes zero $r_h \rightarrow 0$, the quasinormal modes should approach their pure AdS values, i.e., the normal modes of AdS space-time without a black holes, as it was proved in \cite{Konoplya:2002zu} for the scalar field and in
\cite{Natario:2004jd} for the gravitational field in four spacetime dimensions.  In our higher-dimensional case, we have also checked this to confirm that in the limit $a = 0$ and $r_h \rightarrow 0$, the quasinormal frequencies approach their normal AdS values for tensor-type gravitational perturbations
\begin{equation}
\omega_{n} R \rightarrow 2 n + D + 2 l + m + j - 1, \quad r_h \rightarrow 0.
\end{equation}
For non-zero rotations, the quasinormal frequency in the limit $r_h \rightarrow 0$ is characterized by that in the pure AdS case again, but with a finite correction $F$ depending on $a$:
\begin{equation}
\omega_{n} R \rightarrow 2 n + D + 2 l + m + j - 1 + F(D, l, m, j, a), \quad r_h \rightarrow 0.
\end{equation}
We can see representative examples of the dependence of the first few QNMs on $a$ for a fixed $r_{h}$ in Fig. \ref{fig:overtones}. In this figure, we also observe that higher overtones are naturally more stable, as they damp quicker.

One should note that here we have considered the limit of $r_h \rightarrow 0$ keeping $R$ constant ($R=1$).  This is different from another small black hole limit for which $R\rightarrow\infty$ keeping $r_h$ and $a$ finite. The latter limit corresponds to Kerr black holes in the asymptotically flat background.

%T3>fig:m-dep
\begin{figure}
\ifdviout
\includegraphics[width=0.5\linewidth]{\FigDir/KerrAdS_fundamental_r0=0_5_Re-1.eps},\includegraphics[width=0.5\linewidth]{\FigDir/KerrAdS_fundamental_r0=0_5_Im-1.eps}
\else
\includegraphics[width=0.5\linewidth]{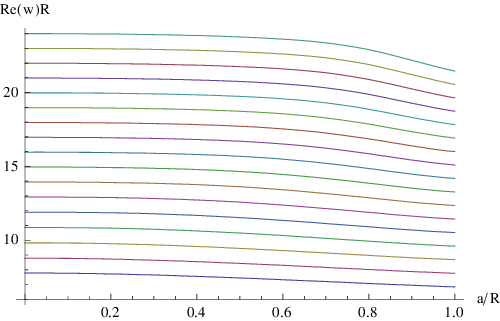},\includegraphics[width=0.5 \linewidth]{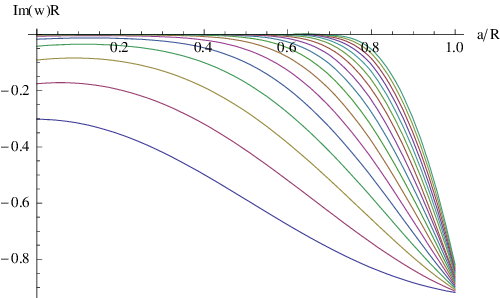}
\fi
\caption{Real (left) and imaginary (right) parts of the fundamental modes, for $r_h = 0.5$ as a function of $a$, $m = 0, 1, ...16$, $D=7$.}\label{fig:m-dep}
\end{figure}

Before going to the arguments on instability, let us discuss the dependence of quasinormal modes on the parameters $D$, $m$, $n$, $r_h$ and $a$.
Figure \ref{fig:m-dep} shows representative examples of the dependence of the fundamental modes on $m$.
One can see that the spacing in the frequency $\omega$ between adjacent $m$-modes  at large $m$ becomes independent of $m$ as
\begin{equation}\label{highm}
\Re(\omega_{m+1}) - \Re(\omega_{m}) = R^{-1}+{\cal O}\left(\frac{1}{m}\right), \quad \Im (\omega_{m+1}) - \Im(\omega_{m}) = {\cal O}\left(\frac{1}{m}\right)<0.
\end{equation}
Thus, a mode with higher $m$ has the negative imaginary part with a larger absolute value, so is more stable. This is rather unexpected because the increase of $m$ makes the negative dip of the potential deeper as can be seen from \eqref{EffectivePot:radial}. One possible explanation is that  because the eigenvalue $\mu$ is approximately given by $m^2$ at large $m$ as \eqref{mu:pureAdS} suggests, the increase of $m$  also increases the height of potential barrier, which supercedes the former effect.

%T3>fig:super_m
\begin{figure}
\ifdviout
\includegraphics[width= 0.49 \linewidth]{\FigDir/KerrAdS_fundamental_r0=10_Re.eps}
\includegraphics[width= 0.49\linewidth]{\FigDir/KerrAdS_fundamental_r0=10_Im.eps}
\else
\includegraphics[width= 0.49 \linewidth]{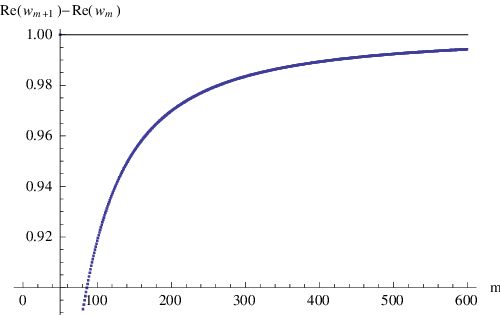}
\includegraphics[width= 0.49\linewidth]{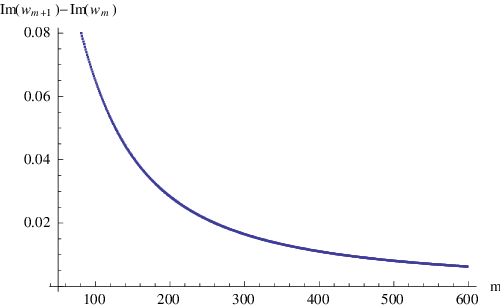}
\fi
\caption{Spacing between real (left) and imaginary (right) parts of high $m$-modes ($D=7$, $r_h=10$, $R=1$). In the right panel, $\Im(\omega_{m+1})-\Im(\omega_m)$ is actually negative and its absolute value is plotted. }\label{fig:super_m}
\end{figure}

In order to check the formula (\ref{highm}) for other values of the background parameters, we have considered a large black hole $r_h=10R$ and calculated $\omega_m$ for $m = 1,2, 3,...600$. As shown in Fig. \ref{fig:super_m}, the high-$m$ spacing behaves in the same way as above in the units $R=1$. This is qualitatively similar to the high-multipole frequency behavior for $4$-dimensional Schwarzschild black holes, which is known analytically. These results suggest that the formula (\ref{highm}) holds for all values of $D$, $r_h$ and $a$.

%T2>Instability
\subsection{Instability}

%T3>fig:Z_n
\begin{figure}
\ifdviout
\includegraphics[width=0.5 \linewidth]{\FigDir/KerrAdS_m=10_r0=0_5_a=0_8_Abs.eps},\includegraphics[width=0.5\linewidth]{\FigDir/KerrAdS_m=10_r0=0_5_a=-0_8_Abs.eps}
\else
\includegraphics[width=0.5 \linewidth]{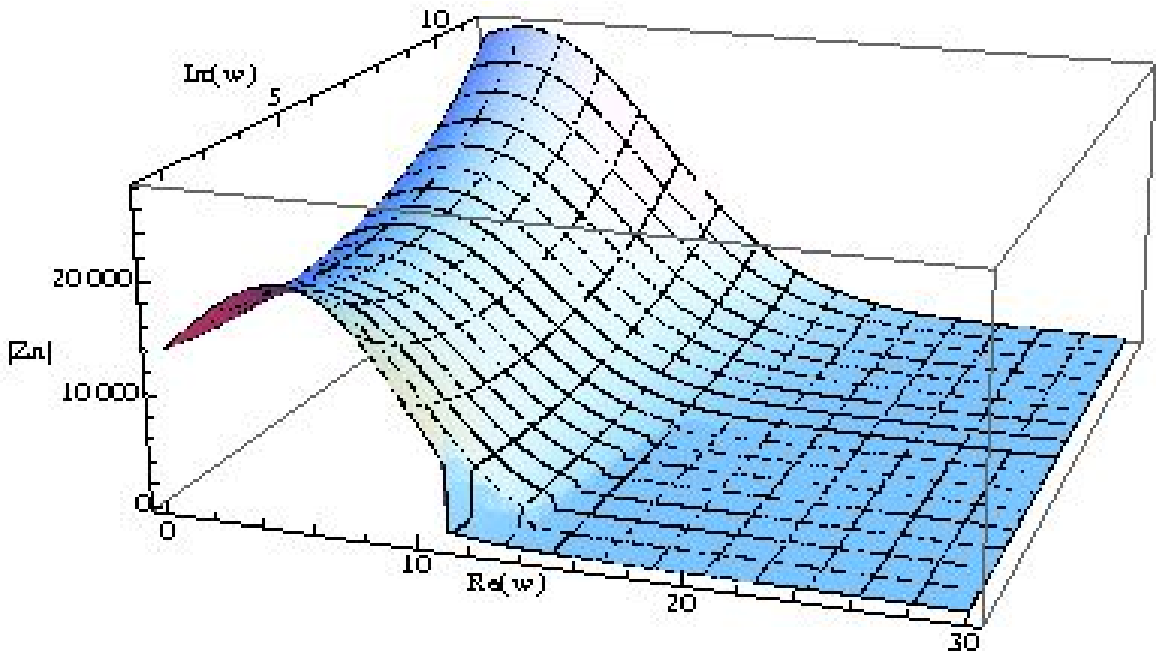},\includegraphics[width=0.5 \linewidth]{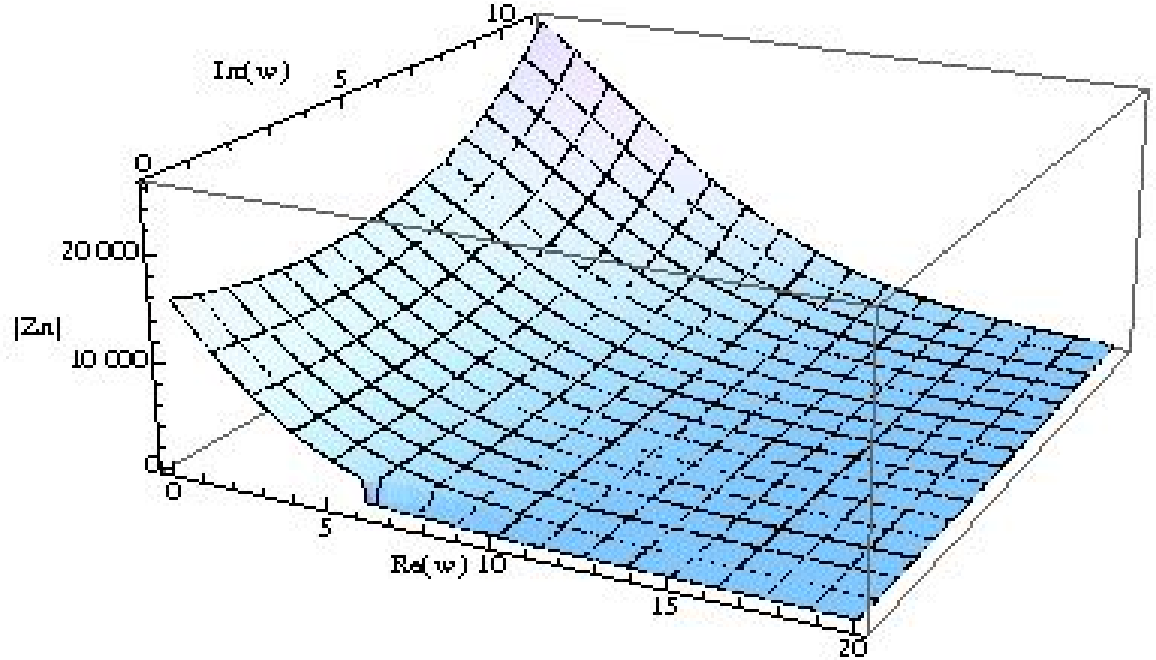}
\fi
\caption{Absolute values of $Z_n(\omega)$ for $m = 10$, $r_h = 0.5$, $a = 0.8$ (left) and $a=-0.8$ (right).}\label{fig:Z_n}
\end{figure}

%T3>fig:D=7,rh=0.1,m=2,3
\begin{figure}
\ifdviout
\includegraphics[width= 0.5 \linewidth]{\FigDir/KerrAdS_fundamental_m=2_r0=0_1.eps},\includegraphics[width= 0.5\linewidth]{\FigDir/KerrAdS_fundamental_m=3_r0=0_1.eps}
\else
\includegraphics[width= 0.5 \linewidth]{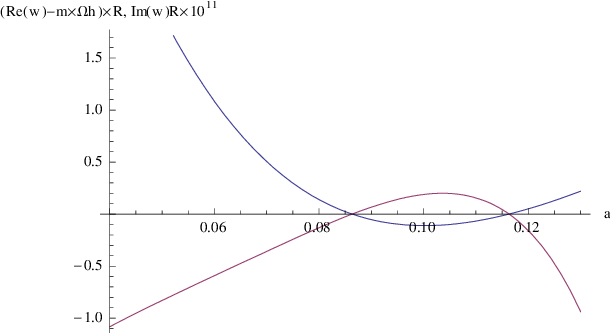},\includegraphics[width= 0.5 \linewidth]{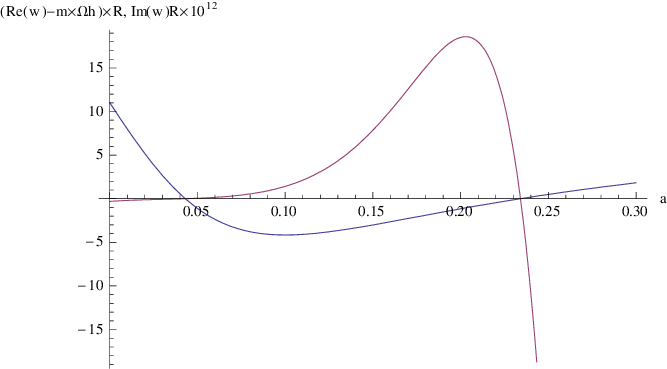}
\fi
\caption{Fundamental quasinormal modes $m=2$(left) and $m=3$(right), $r_h = 0.1$ as a function of $a$ ($D=7$). Blue line is the $R (\Re(\omega) - m \Omega_h) $, red line is $\propto R Im(\omega)$.}\label{fig:omega-a:D=7,rh=0.1,m=2,3}%super0
\end{figure}

%T3>fig:D=7,rh=0.1,m=4,5
\begin{figure}
\ifdviout
\includegraphics[width= 0.5 \linewidth]{\FigDir/KerrAdS_fundamental_m=4_r0=0_1.eps},\includegraphics[width= 0.5 \linewidth]{\FigDir/KerrAdS_fundamental_m=5_r0=0_1.eps}
\else
\includegraphics[width= 0.5 \linewidth]{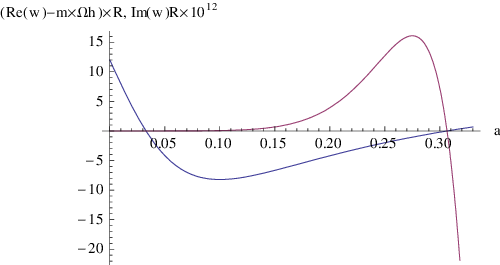},\includegraphics[width= 0.5 \linewidth]{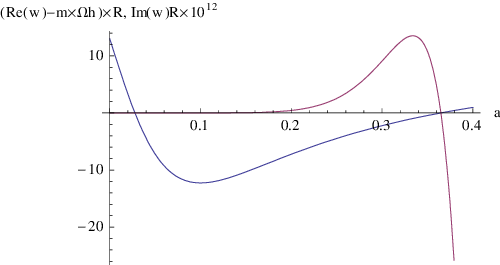}
\fi
\caption{Fundamental quasinormal modes $m=4$(left) and $m=5$(right), $r_h = 0.1$ as a function of $a$ ($D=7$). Blue line is the $R (\Re (\omega) - m \Omega_h) $, red line is $\propto R Im (\omega)$.}\label{fig:omega-a:D=7,rh=0.1,m=4,5}%super1
\end{figure}

%T3>fig:D-8,rh=0.1,m=2,3
\begin{figure}
\ifdviout
\includegraphics[width= 0.49 \linewidth]{\FigDir/KerrAdS_fundamental_n=4_m=2_r0=0_1.eps}
\includegraphics[width= 0.49 \linewidth]{\FigDir/KerrAdS_fundamental_n=4_m=3_r0=0_1.eps}
\else
\includegraphics[width= 0.49 \linewidth]{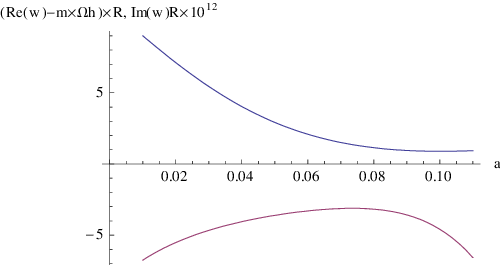}
\includegraphics[width= 0.49 \linewidth]{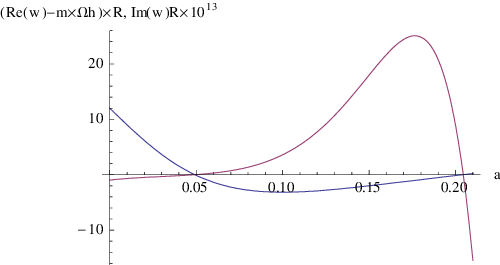}
\fi
\caption{Fundamental quasinormal modes $m=2$(left) and $m=3$(right),  $D=8$, $r_h = 0.1$ as a function of $a$. Blue line is the $R (\Re (\omega) - m \Omega_h) $, red line is $\propto R Im (\omega)$.}\label{fig:omega-a:D=8,rh=0.1,m=2,3}%super2
\end{figure}

%T3>fig:D=8,rh=0.1,m=4,5
\begin{figure}
\ifdviout
\includegraphics[width= 0.49 \linewidth]{\FigDir/KerrAdS_fundamental_n=4_m=4_r0=0_1.eps}
\includegraphics[width= 0.49 \linewidth]{\FigDir/KerrAdS_fundamental_n=4_m=5_r0=0_1.eps}
\else
\includegraphics[width= 0.49 \linewidth]{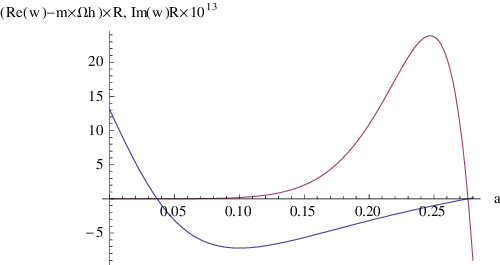}
\includegraphics[width= 0.49 \linewidth]{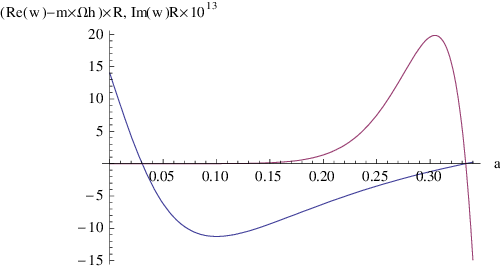}
\fi
\caption{Fundamental quasinormal modes $m=4$(left) and $m=5$(right),  $D=8$, $r_h = 0.1$ as a function of $a$. Blue line is the $R (\Re (\omega) - m \Omega_h) $, red line is $\propto R Im (\omega)$.}\label{fig:omega-a:D=8,rh=0.1,m=4,5}%super3
\end{figure}

%T3>fig:D=10,rh=0.1,m=5
\begin{figure}
\ifdviout
\includegraphics[width= 0.49\linewidth]{\FigDir/KerrAdS_fundamental_n=6_m=5_r0=0_1.eps}
\includegraphics[width= 0.49\linewidth]{\FigDir/KerrAdS_fundamental_n=3_m=2_r0=0_02.eps}
\else
\includegraphics[width= 0.49 \linewidth]{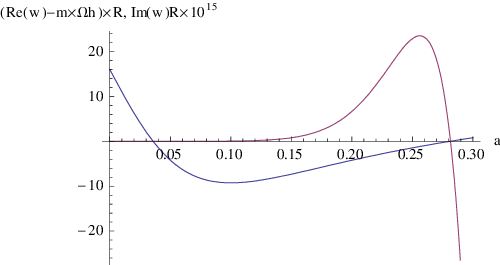}
\includegraphics[width= 0.49\linewidth]{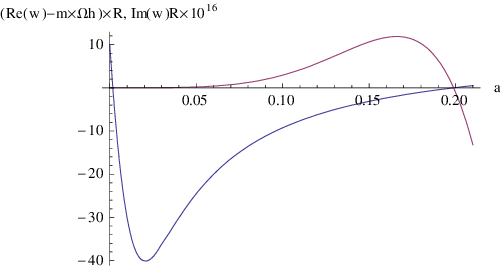}
\fi
\caption{Fundamental quasinormal modes $m=5$,  $D=10$, $r_h = 0.1$(left) and $D=7$, $m=2$, $r_h =0.02$(right) as a function of $a$. Blue line is the $R (\Re (\omega) - m \Omega_h) $, red line is $\propto R \Im (\omega)$.}\label{fig:omega-a:D=10,rh=0.1,m=5}%super_D81
\end{figure}

The first step in the search for unstable modes is to check that there occurs no instability in the non-superradiant range of the spectrum. For this, we have performed extensive calculations of $|Z_n(\omega)|$ in a wide range of complex values for $\omega$ with a positive imaginary part, which corresponds to instability, and found no zeros of $|Z_n(\omega)|$  if the superradiance condition
\begin{equation}\label{superradiance}
\Re (\omega) < m \Omega_h
\end{equation}
is not satisfied (see  Fig. \ref{fig:Z_n} for representative examples).

In order to get an initial hint on where to look for superradiant unstable modes, as the next step,  we used the following estimation. Since the suspected region of instability is $1 > a > r_h^2$ in units $R=1$, we are dealing with small black holes including those for which the QNMs are close to the pure AdS normal modes, that is, $\Re (\omega) \approx 2 n + D + 2 l + m + j - 1$. For these black holes, the superradiance inequality $\Re (\omega) < m \Omega_{h}$ leads to the inequality $2 n + D + 2 l + m + j - 1 < m \Omega_{h}$, which gives us an initial guess in the numerical search for unstable modes with small $a$. In particular, from this inequality and $\Omega_h=a(1+r_h^2)/(r_h^2+a^2)>1$, we find that the smaller the black hole is, the smaller the minimal value of $m$ for superradiance is. Thus, for instance for  $D=7$, $r_h =0.5$, $l =2$ and $j=0$, the superradiance occurs only for $m \geq 32$, while for $r_h =0.1$ and the same  $D, l$ and $j$, the superradiance starts already at $m \geq 2$. The above inequality also implies that for higher $D$ and the same fixed $r_h$, the minimal value of $m$ for the superradiant instability slowly increases with $D$. Thus, for example,  for $D=8$, $r_h =0.1$, $l =2$ and $j=0$, the minimal azimuthal number for the instability is $m=3$, while $m=2$ is not in the superradiant regime (see Fig. \ref{fig:omega-a:D=8,rh=0.1,m=2,3})

Now, we summarize the basic features of the results of our numerical calculations.
\Bitm
\item[1)] The $\Im \omega$ is positive \emph{(unstable) only for modes in the superradiant regime} (see Figs. \ref{fig:omega-a:D=7,rh=0.1,m=2,3}-\ref{fig:omega-a:D=10,rh=0.1,m=5}).
\item[2)] The instability growth rate is tiny and of order $10^{-12}$ or less in units $R=1$.
\item[3)] The maximal value of the instability growth rate, $\Gamma$, decreases when $D$ increases, at least in the calculated cases with $D$ in the range $7\le D\le 10$. For example,  for $D=7$, $r_h = 0.1$ and $a \approx 0.21$ in units  $R=1$, $\Gamma= 1.8 \times 10^{-11}$, while for higher $D$, the instability growth is smaller (see Fig. \ref{fig:omega-a:D=8,rh=0.1,m=2,3}, \ref{fig:omega-a:D=8,rh=0.1,m=4,5}, \ref{fig:omega-a:D=10,rh=0.1,m=5}).
\item[4)] The unstable range of $a$ increases when we increase $m$. Black holes becomes unstable for slower rotations with the increase of the azimuthal number $m$.
\Eitm

The last observation together with the numerical formula (\ref{highm}) allows us to guess an analytic expression for the region of instability. In fact, the superradiance inequality (\ref{superradiance}) for large $m$ reads
$$
\frac{\Re(\omega)}{m}=\frac{1}{R}+{\cal O}\left(\frac{1}{m}\right)<\Omega_h.
$$
Since the instability region in the parameter space expands with the increase of  $m$, the $m\rightarrow\infty$ limit gives us the \emph{exact formula} for the unstable region
\Eq{
\Omega_h R>1.
}
For any values of $a$, $r_h$ and $R$ in this region, we can find a value of $m$ that satisfies the superradiance inequality, and therefore, the corresponding perturbations are unstable.

Taking into account the regularity condition $a<R$ and the equation \eqref{ROmega}, we conclude that the parameter region of instability for rotating AdS black holes against tensor-type gravitational perturbations is given by the inequality
\Eq{
r_h^2<a R<R^2.
}
This corresponds to the white (top-left) region on Fig. \ref{fig:paramregion}.

The tiny instability growth takes place also for a massive scalar field around an asymptotically flat Kerr black hole. Thus, it is probable that a small growth rate is generic, although it is a poorly-understood feature of the superradiant instability of black holes. Note that such a tiny classical instability of small asymptotically AdS black holes will not probably affect the dynamic of small black holes, for which the quantum thermodynamic instability and Hawking evaporation are much more effective and violent processes. Moreover, even pure classically, the superradiant instability of a black hole in the AdS box does not necessarily mean that the black hole disappears or collapses to singularity.  At least near the border of the unstable parameter region (the white top-left region on Fig. \ref{fig:paramregion}), it would be natural for a black hole to diminish  its angular momentum until it reaches a stable state, which is still described by a Myers-Perry-AdS solution with smaller $a$. The complete and definite answer on the evolution and destiny of rotating black holes in AdS space should be given by full non-linear analysis of perturbations.

%T1>Conclusion
\section{Conclusion}

In the present paper, we have derived the master equation for the tensor-type gravitational perturbations of a simply rotating Myers-Perry black hole with the $\Lambda$-term in the spacetime dimension $D\geq7$. We have analytically proved the stability of black holes obeying inequality  $a\le r_h^2/R$, while for small rotating AdS black holes violating this condition, 
we have found unstable modes. We further found that the parameter range for instability exactly coincides with that for superradiance instability. This result indicates that one should be careful when one utilizes a stationarity of the black hole metric as an implicit background for propagation of fields, for example when one calculates the grey-body factors (which implicitly provide the emission probabilities for the Hawking radiation) or quasinormal modes of higher dimensional black holes \cite{Kanti:2004nr}.

We have also found a strong numerical evidence that the parameter region for unstable black holes is identical to the region $R\Omega_h>1$ or equivalently $a> r_h^2/R$ in which the analytical stability proof does not hold. In addition, we have analyzed the spectrum of the damped quasinormal modes in the stable sector and found some general tendency in the dependence of the complex frequency on the mode parameters and the black hole parameters.

Although we have found some parameter region of instability for rotating AdS black holes here, it is possible that the complete parameter region of instability is larger than the one found here, because we have considered only tensor-type gravitational perturbations. In particular, our analysis is powerless in studying the stability of rotating black holes in five or six dimensions. In order to investigate the stability of these black holes, it is clear that the scalar-type and vector-type gravitational perturbations have to be studied. Unfortunately, these are at present out of our reach because perturbation equations in those cases have not been reduced to a set of tractable ODEs by the separation of variables except for some exceptional cohomogeneity one cases such as the special five-dimensional Myers-Perry black holes with $a_1=a_2$ \cite{Murata:2008yx}.

In spite of this restriction, there still exist a lot of interesting problems that can be handled by the present formulation. Some examples are the investigations of the tensor-type gravitational quasinormal modes for simply rotating black holes in the asymptotically flat case and in the asymptotically de Sitter case. The above formulation can be also used to calculate the classical scattering amplitudes as well as the probability of emissions (due-to the Hawking radiation) of tensor-type gravitons from a simply rotating black hole.

%T1>Acknowledgements

%\newpageen
\begin{acknowledgments}
R. A. K. was supported by \emph{the Japan Society for the Promotion of Science}, Japan. R. A. K also acknowledges the hospitality of the Theory Division of the High Energy Accelerator Research Organization (KEK) in Tsukuba, where the final part of this work was done.
A. Z. was supported by \emph{Funda\c{c}\~ao de Amparo \`a Pesquisa do Estado de S\~ao Paulo (FAPESP)}, Brazil.  H. K. was supported in part by Grants-in-Aid for Scientific Research from JSPS (No. 18540265).
\end{acknowledgments}

%T1>References
%\bibliographystyle{jphys_n_usrt}
%\bibliography{\BibFile}

\end{document}

%T1>EOF
\end{document}